\documentclass{article}
\usepackage[utf8]{inputenc}
\usepackage{authblk}
\usepackage[british]{babel}
\usepackage{amsmath,graphicx,amssymb, amsfonts, amsthm}
\usepackage{subfigure}

\DeclareMathOperator*{\argmin}{arg\,min}

\title{Kam method for Cryo-EM particle reconstruction}
\author{Yin Xian}
\affil{Department of Mathematics, Hong Kong University of Science and Technology, Clear Water Bay, Hong Kong}

\date{September 2018}

\begin{document}

\maketitle

\section*{Abstract}
The Cryo-EM 3D particle reconstruction is essential for  identifying protein and uncover the biological mechanism of the macro-molecules. In this paper, we use Kam method for reconstruction. Kam method is \textit{ab-initio}, and it assumes that the projection angles of the particle are uniformly distributed. Based on the data covariance matrix, we compute the radial frequency component of the matrix. The particle density function can be obtained by the radial frequency component and the angular frequency basis function. In order to uniquely and accurately identify the radial frequency component, an initial guess of the structure is applied. Experiment shows that Kam method works for low resolution particle reconstruction. 
Further improvement can be made by including projection angle distribution in covariance matrix, and applying  the fast algorithm to enhance computation speed.

\section{Introduction}
Single Particle Reconstruction (SPR) from Cryo-Em is an increasingly popular technique in structural biology for determining 3D structures of macro-molecular complexes. The challenge of Cryo-EM particle reconstruction is that the viewing directions of the images are unknown. The noise in the Cryo-EM data is also heavy. 

The imaging process of Cryo-EM is as follows. 
\begin{align}
    I(x,y)=p*\int_{-\infty}^{\infty}\phi(R^Tr)dz+n
\end{align}
where $\phi$ is the particle density, $I$ is the projected image, $R$ is the rotation, $p$ is the point spread function of the microscope, $n$ is the noise, and $r=(x,y,z)$. Current techniques to reduce the noise of the Cryo-EM image can be found in the literature~\cite{xian2018}. 
The 3D reconstruction problem is challenge and it is a nonlinear inverse problem. 

For 3D particle reconstruction, it is typical to guess an initial structure and perform an iterative refinement procedure. When the image is noisy, the refinement process depends heavily on the initial model. The method of moments~\cite{goncharov1988, salzman1990} and the common-lines based models~\cite{vanheel1987} are two known approaches for \textit{ab-initio} estimation. For common-lines based approaches, it was able to obtain \textit{ab-initio} reconstruction from real microscope images. However, these algorithms are sensitive to noise, and so far not so successful in obtaining meaningful 3D \textit{ab-initio} models directly from raw images. 

The Kam method requires that the number of the collected images is large enough for accurate estimation of the covariance matrix of the 2D projection images. In this paper, we test the efficiency of reconstructing the Cryo-EM 3D macro-molecule, and seek to improve the Kam method for reconstruction. 

\section{Kam method}
In Cryo-Electron Microscopy (Cryo-EM), the 3D structure of a molecule needs to be determined from its 2D projection images taken at unknown viewing directions. 

Kam~\cite{kam1977, kam1980} showed that the autocorrelation function of the 3D molecule over the rotation group SO(3) can be estimated from 2D projection images whose viewing directions are uniformly distributed over the sphere. Let $\phi$ be the 3D density function, $\hat{\phi}$ be the 3D Fourier transform of $\phi$. Its expansion in spherical coordinates is:
\begin{align}
\hat{\phi}(k,\theta,\phi)=\sum\limits_{l=0}^{\infty}\sum\limits_{m=-l}^{l} A_{lm}(k)Y_l^m(\theta,\phi)
\end{align}
where $k$ is the radial frequency and $Y_l^m$ are the real spherical harmonics. Kam show that 
\begin{align}
C_l(k_1,k_2)=\sum\limits_{m=-l}^{l} A_{lm}(k_1)\overline{A_{lm}(k_2)}
\label{eq:covariance}
\end{align}
can be estimated from the covariance matrix of the 2D projection images. 
For image sampled on a Cartesian grid, $C_l$ is of size $K\times K$, where $K$ is the maximum frequency.

The formulation is shown as follows. The scattering intensity pattern from one of the particles as a function of the scattering vector $\kappa$ is given by (square of Fourier transform, amplitude square)
\begin{align*}
S(\omega, \kappa)=\left|\int dr \rho(R(\omega)r)e^{i\kappa\cdot r}\right|^2=|A(\omega,\kappa)|^2
\end{align*}
where $R(\omega)$ is the rotation operator that rotates the particle, $\rho(r)$ is the initial position. $\omega$ defines the three Eulerian angles of rotation of a rigid body $\omega=(\alpha, \beta, \gamma)$. 

\begin{figure}[!ht]
\centering
{\includegraphics[width=.45\linewidth,height=5cm]{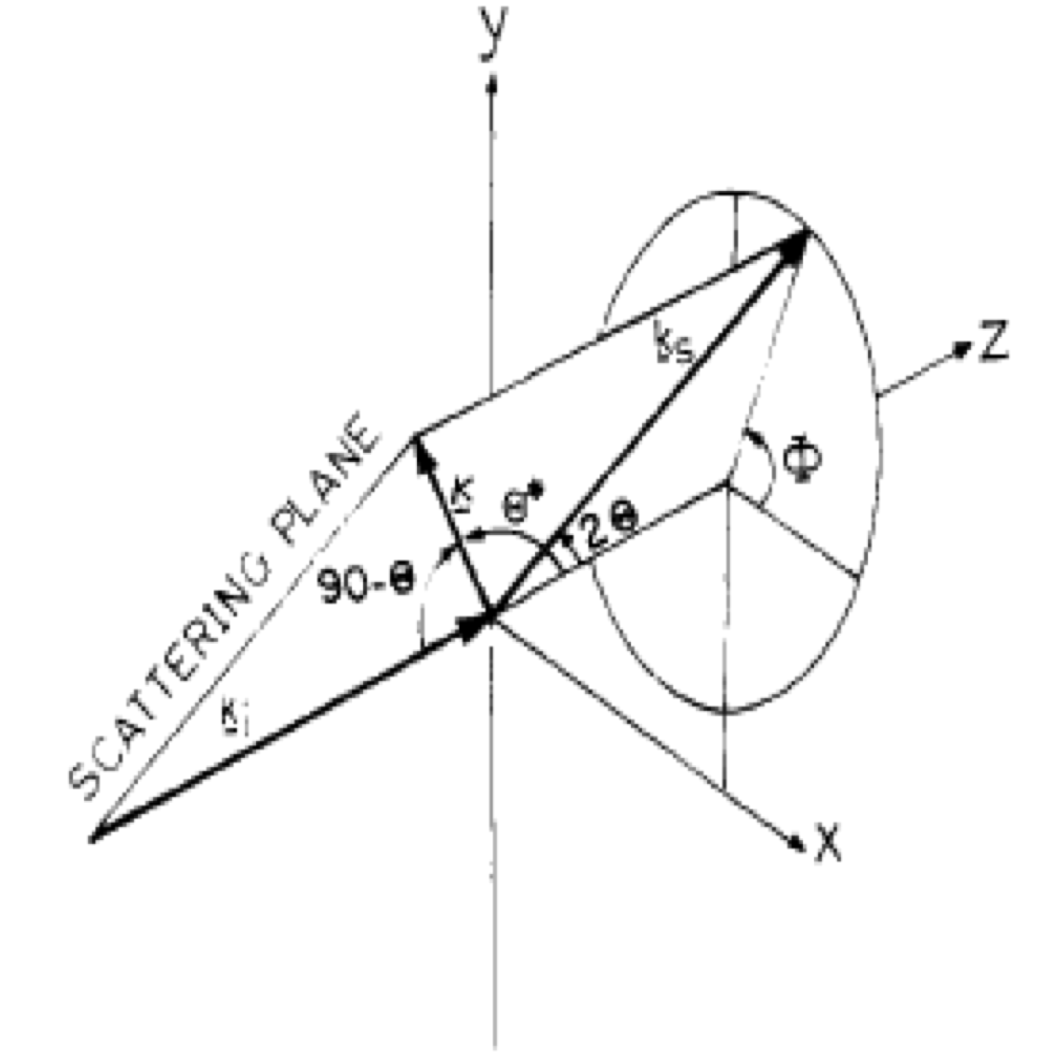}}
\caption{Illustration of the scattering configuration. The $z$ axis was taken along the incident beam vector $k_{i}$. The polar angles $(2\theta, \phi)$ of the scattered beam vector $k_s$ and $(\theta^{*}, \phi)$ of the scattering vector $\kappa=k_s-k_i$ are marked. }
\label{fig:class}
\end{figure}

The scattering vector $\kappa$ is defines by the incident vector $k_i$ and the scattered vector $k_s$. 
\begin{align*}
\kappa=k_s-k_i
\end{align*}

Its magnitude is 
\begin{align*}
|\kappa|=2|k_i|\sin\theta=2|k_i||\cos\theta^{*}|=2|k_s|\sin\theta
\end{align*}
The scattered photon direction $k_s$ is defined by the angles $\Omega=(2\theta,\Phi)$, and the direction of $\kappa$ by $\Omega^{*}$

Averaging over the particle orientation, 
\begin{align*}
S(\kappa)=\frac{N}{8\pi^2}\int S(\omega, \kappa) d\omega=S(|\kappa|)
\end{align*}
It is axially symmetric around the $z$ axis, and 
\begin{align*}
\int d\omega=\int_{-\pi}^{\pi} d\gamma\int_{0}^{\pi} \sin\beta d\beta\int_{0}^{2\pi} d\alpha=8\pi^2
\end{align*}

Define the spatial correlation function as,
\begin{align}
C(\kappa_1,\kappa_2)=\frac{N}{8\pi^2}\int S(\omega,\kappa_1) S(\omega, \kappa_2) d\omega
\label{eq:cov_3d}
\end{align}

Considering the rotation of the particle, the spherical harmonic expansion of $S(\omega, \kappa)$ is 
\begin{align*}
S(\omega, \kappa)=\sum\limits_{lmm'}A_{lm}(|\kappa|)Y_{lm'}(\Omega^{*})R_{mm'}^{l}(\omega)
\end{align*}

There's orthogonality property of the rotation matrix 
\begin{align*}
\int R_{m'm}^{l}(\omega)R_{M'M}^{l}(\omega)d\omega = \delta_{l'l}\delta_{mM}\delta_{m'M'}\frac{8\pi^2}{2l+1}
\end{align*}
also,
\begin{align*}
\sum\limits_{m'}Y_{lm'}(\Omega_1^{*})Y_{lm'}(\Omega_2^{*})=\frac{2l+1}{4\pi}P_l(\cos\Psi)
\end{align*}
where $\Psi$ is the angle between the direction $\Omega_1^*$ and $\Omega_2^*$ of $\kappa_1$ and $\kappa_2$. Substitute the above formula to eq.~(\ref{eq:cov_3d}), we have (angle between two vector)
\begin{align*}
C(\kappa_1, \kappa_2)=\frac{N}{4\pi}\sum\limits_{l} P_l(\cos\Psi)\left[ \sum\limits_{m=-l}^{l}A_{lm}(|\kappa_1|)A_{lm}^{*}(|\kappa_2|)\right]= C(|\kappa_1|, |\kappa_2|, \Psi)
\end{align*}

Using the orthogonality of the Legendre polynomials $P_l(\cos\Psi)$, decompose $C$ into $l$ different subspaces, we have 
\begin{align*}
C_l(|\kappa_1|,|\kappa_2|)&=\frac{2\pi(2l+1)}{N}\int_{0}^{\pi}C(\kappa_1,\kappa_2)P_l(\cos\Psi)\sin\Psi d\Psi \\
&=\sum\limits_{m=-l}^{l}A_{lm}(|\kappa_1|)A_{lm}^{*}(|\kappa_2|)
\end{align*}

The use of autocorrelation is also frequent in acoustic signal processing community~\cite{xian2017,xian2015}. 

\section{Spherical harmonic decomposition}
If $V$ is the space of functions on the sphere, we can consider the sub-space of functions on the sphere that are restrictions of homogeneous polynomials of degree $d$.  

Since a rotation will map a homogeneous polynomial of degree $d$ back to a homogeneous polynomial of degree $d$, these sub-spaces are sub-representations. If $(x,y,z)$ is a point on the unit sphere, and it satisfies $x^2+y^2+z^2=1$. Thus, if $q(x,y,z)\in P_d(x,y,z)$, the polynomial $q(x,y,z)(x^2+y^2+z^2)$ is a polynomial of degree $d+2$, its restriction to the sphere is actually a homogeneous polynomial of degree $d$. While the sub-spaces $P_d(x,y,z)$ are sub-representations, they are not irreducible as $P_{d-2}(x,y,z)\subset P_d(x,y,z)$. To get the irreducible sub-representations, look at the spaces
$V_d=P_d(x,y,z)\cap P_{d-2}(x,y,z)^{\perp}$, and the dimension of these sub-representation is 
\begin{align*}
\dim(V_d)&=\dim(P_d(x,y,z))-\dim(P_{d-2}(x,y,z)) \\
&=\sum\limits_{i=0}^{d}(i+1)-\sum\limits_{i=0}^{d-2}(i+1) \\
&=2d+1
\end{align*}

The spherical harmonics of frequency $d$ are an orthonormal basis for the space of functions $V_d$. If we represent a point on a sphere in terms of its angle of elevation and azimuth,
\begin{align*}
(\theta, \phi)=(\cos\phi\sin\theta, \cos\theta, \sin\phi\sin\theta)
\end{align*}
with $0\leq \theta\leq \pi$ and $0\leq\phi< 2\pi$.

The spherical harmonics are functions $Y_l^m$, with $l$ and $m$ are integers, and $l\geq 0$ and $|m|\leq l$, spanning the sub-representations $V_l$.
\begin{align*}
V_0&=Span\{Y_0^0(\theta, \phi)\} \\
V_1&=Span\{Y_1^{-1}(\theta, \phi), Y_1^0(\theta, \phi), Y_k^k(\theta, \phi)\} \\
&\cdots \\
V_k&=Span\{Y_k^{-k}(\theta, \phi), Y_k^{-k+1}(\theta, \phi), \cdots, Y_k^{k}(\theta, \phi)\} 
\end{align*}

If we have a function defined on a sphere, sample on a regular $n\times n$ grid of angles of elevation and azimuth, the forward and inverse spherical harmonic transforms can be computed in $O(n^2\log^{2}n)$. Like the FFT, the fast spherical harmonic transform can be thought of as a change of basis, and a brute force method would take $O(n^4)$ time.

For spherical coordinates, the angular part of a basis function is a spherical harmonic
\begin{align*}
Y_{l}^m(\theta,\phi)=\sqrt{\frac{2l+1}{4\pi}\frac{(l-m)!}{(l+m)!}}P_l^m(\cos\theta)e^{im\phi}
\end{align*}
$P_l^m$ is the associated Legendre polynomial, where
\begin{align*}
P_l^m(z)=(-1)^m(1-z^2)^{m/2}\frac{d^m}{dz^m}P_l(z)
\end{align*}
$P_l$ is the Legendre polynomials
\begin{align*}
P_l(z)=\frac{1}{2\pi i}\oint(1-2tz+t^2)^{1/2}t^{-n-1}dt
\end{align*}

The $Y_l^m$ are the eigen-values of the Laplacian operator
\begin{align*}
\nabla^2 f(\theta,\phi)=\lambda f(\theta,\phi)
\end{align*}

The importance of the spherical harmonics is that they are an orthonormal basis for the $(2d+1)$ dimensional sub-representations. 

\subsection{Sub-Representation}
The $Y_l^m$ are spherical functions whose number of lobes get larger as the frequency $l$ gets bigger. 
\begin{figure}[!ht]
\centering
{\includegraphics[width=.6\linewidth,height=3.5cm]{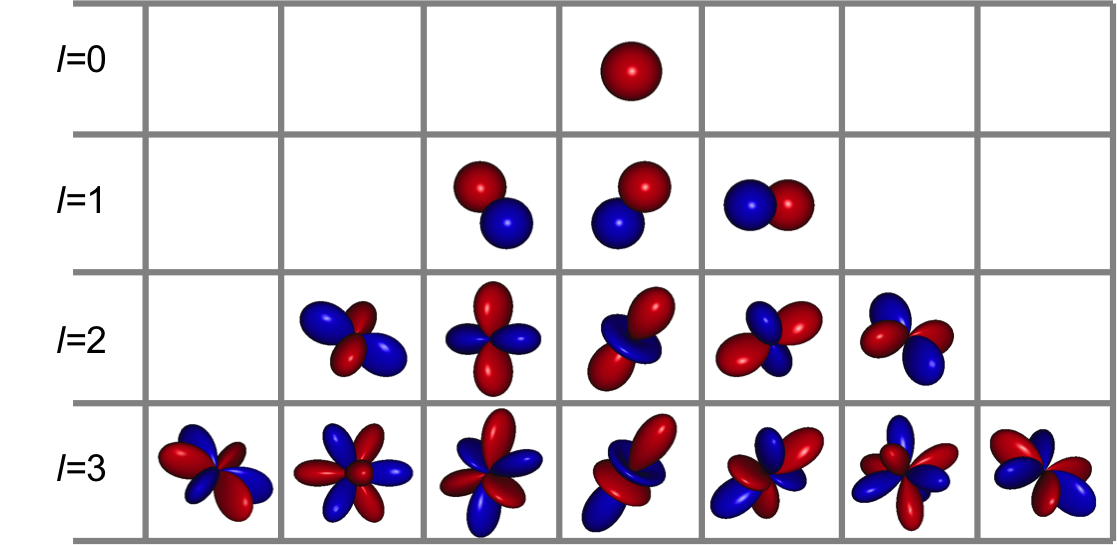}}
\caption{Sub-representation of spherical harmonics}
\label{fig:class}
\end{figure}

\subsection{Rotation invariance}
Given a spherical function $f$, we can obtain a rotation invariance representation by expressing $f$ in terms of its spherical harmonic decomposition
\begin{align*}
f(\theta, \phi)=\sum_{l=0}f_l(\theta, \phi)
\end{align*}
where each $f_l\in V_l$
\begin{align*}
f_l(\theta,\phi)=\sum\limits_{m=-l}^{l}a_l^mY_l^m(\theta,\phi)
\end{align*}

We can obtain a rotational invariant representation by storing the size of each $f_l$ independently
\begin{align*}
\Psi(f)=\{||f_0||, ||f_1||, \cdots, ||f_l||, \cdots\}
\end{align*}
where
\begin{align*}
||f_l||=\sqrt{||a_l^{-l}||^2+||a_l^{-l+1}||^2+\cdots+||a_l^{l-1}||^2+||a_l^{l}||^2}
\end{align*}
By storing only the energy in the different frequencies, we discard information that does not depend on the pose of the model.

\subsection{Correlation}
Given two spherical functions $f$ and $g$, to compute the distance between $f$ and $g$ at every rotation, the correlation
\begin{align*}
Corr<f,g,R>=<f,R(g)>
\end{align*}
need to be computed at every rotation $R$. In spherical representation,
\begin{align*}
f(\theta, \phi)&=\sum\limits_{l=0}\sum\limits_{m=-l}^{l}a_l^mY_l^m(\theta,\phi) \\
g(\theta, \phi)&=\sum\limits_{l=0}\sum\limits_{m=-l}^{l}b_l^mY_l^m(\theta,\phi) \\
<f,R(g)>&=\bigl<\sum\limits_{l=0}\sum\limits_{m=-l}^{l}a_l^mY_l^m, R(\sum\limits_{l'=0}\sum\limits_{m'=-l'}^{l'}b_{l'}^{m'}Y_{l'}^{m'})\bigr>
\end{align*}

\section{Application on Cryo-EM reconstruction}
Bhamre et.~al~\cite{bhamre2015} proposed the orthogonal matrix retrieval algorithms for reconstruction. It lets $A_l=F_lO_l$. It needs to identify $O_l$, which related to the orientations of particle. The orientation is unknown.  It establishes the connection of phase retrieval and the retrieval of the orthogonal matrix $O_l$ . The comparison of X-ray crystallography and cryo-EM is shown~\cite{bhamre2015}.
\begin{table}[h!]
\small
\caption{Comparison of cryo-EM and X-ray crystallography} 
\label{table:1_psnr}
\vspace{1pt}
\centering 
\begin{tabular}{ccc}
\hline\hline
~ & X-ray crystallography  & cryo-EM \\
\hline
Known & magnitudes $|\hat{\phi}|^2=\hat{\phi}\hat{\phi}^{*}$  & $A_lA_l^{*}$   \\
Unknown &  phases $\frac{\hat{\phi}}{|\phi|}$  &  Orthogonal matrix $O_l$ 
\\ [1ex]
\hline 
\end{tabular}
\end{table}

The orthogonal matrix $O_l$ can be considered as ``phase", and $A_lA_l^{*}$ can be considered as the ``magnitude square". 

In order to uniquely identify the unitary matrix, and specify $A_l$, he proposed the Orthogonal Extension (OE) and Orthogonal Replacement (OR) method. 

\begin{figure}[!ht]
\centering
{\includegraphics[width=.5\linewidth,height=3.5cm]{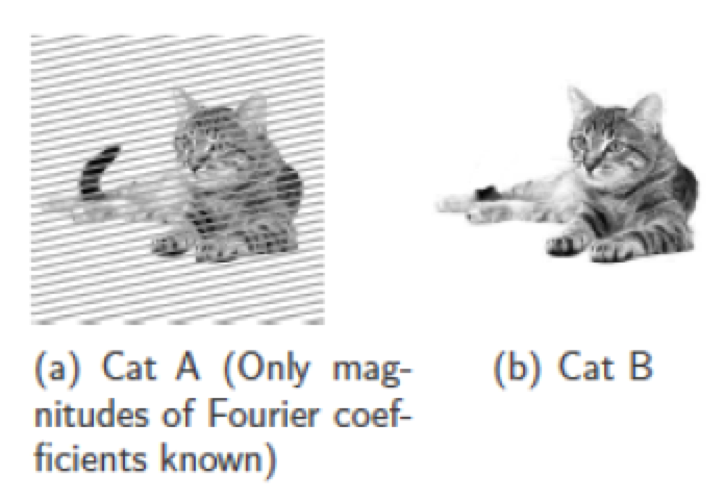}}
\caption{(a) is the truth, (b) is the reference. We try to reconstruct (a) according to the information of (b). Noted that the reference image (b) has no tail.}
\label{fig:cat_1}
\end{figure}

\begin{figure}[!ht]
\center
{\includegraphics[width=.6\linewidth,height=4cm]{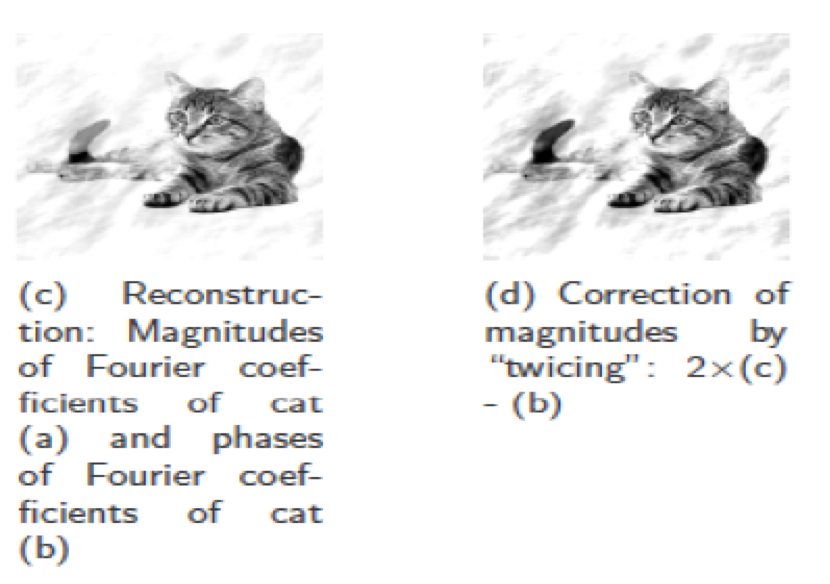}}
\caption{(c) and (d) are both OE solution. (d) applies the ``twicing" scheme. }
\label{fig:cat_2}
\end{figure}

The process of OE is as follows: it estimate the unitary matrix through:
\begin{align}
    O_l=\argmin_{\text{orthogonal matrices } O} ||F_l O-\hat{A}_l||_F^2
\end{align}
where $F_l$ is the spectral solution, $A_l$ is the estimated solution. We obtain $O$ through $A_l$. The solution in this case is: $A_l=F_l O$. In order to enhance the influence of the original information, he proposed the twicing scheme: $A_l=2F_lO-\hat{A}_l$. Illustration of this approach is shown in Figure~\ref{fig:cat_1} and Figure~\ref{fig:cat_2}.

For the OR approach, it introduces more parameters and assumption to obtain the unitary matrix $O$. It seeks to solve the following formula:
\begin{align}
    \min\limits_{O_1,O_2~\text{are orthogonal matrices in }\mathbb{R}^{D\times D}}||F_1O_1+F_2O_2-\hat{A}||_F^2.
\end{align}

The deficiency of the OE and OR approach is that the analogy to phase retrieval is not correct. Phase retrieval cannot reconstruct the global phase. In phase retrieval, more measurement is needed while in Kam's method, only one measurement: the covariance matrix, is obtained. Therefore, the idea of phase retrieval cannot be directly applied in this case. 

\section{Experiment}
\begin{figure}[!ht]
\centering
{\includegraphics[width=.7\linewidth,height=3cm]{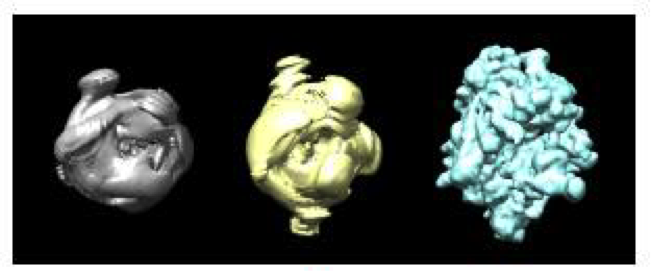}}
\caption{Reconstruction results. The blue one is the result from Relion, the yellow one is the ground truth, and the grey one is the result from Kam method.}
\label{fig:reconstruct}
\end{figure}

We test the algorithm on the EMPIAR-10107 dataset. The data is pre-processed by whiten the noise through the Covariance Wiener Filtering method. It estimates the covariance matrix of the data, and chooses the number of decomposition $L=7$. The total samples of the data is $2\times 10^5$. The result with the use of Kam method for reconstruction is shown in Figure~\ref{fig:reconstruct}. According to the figures, we can see that when the resolution is low, Kam method is more close to the ground truth. The current Kam method needs to be improved in order to process high resolution data. The commercial software, like RELION, may not work well when the data is limited and projection angles of the images are not uniformly distributed.

\section{Summary}
In this paper, we present the Kam method for Cryo-EM 3D particle reconstruction. The Kam method is based on the spherical harmonic decomposition and the autocorrelation of the data. The Kam method intends to reconstruct the particle density function by computing angular frequency basis and the radial frequency component from the covariance matrix. 

The drawback of the algorithm is that the computational cost is high when it computes the covariance matrix of all data, and the results can be further improved by considering the true distribution of projection angles, rather than uniform distribution. Fast algorithm can be applied to speed up the computation.


\begin{thebibliography}{100}
\bibitem{goncharov1988} A. Goncharov, ``Integral geometry and three-dimensional reconstruction of randomly oriented identical particles from their electron microphotos." Acta Applicandae Mathematica \textbf{11.3}, pp: 199-211 (1988).

\bibitem{salzman1990} D. Salzman, ``A method of general moments for orienting 2D projections of unknown 3D objects." Computer vision, graphics, and image processing \textbf{50.2}, pp: 129-156 (1990).

\bibitem{vanheel1987} M. Van Heel. ``Angular reconstitution: a posteriori assignment of projection directions for 3D reconstruction." Ultramicroscopy \textbf{21.2}, pp: 111-123 (1987).



\bibitem{kam1977} Z. Kam. ``Determination of macromolecular structure in solution by spatial correlation of scattering flucturations." Macromolecules, \textbf{82}, pp. 927- 934 (1977).

\bibitem{kam1980} Z. Kam. ``The reconstruction of structure from electron micrographs of randomly oriented particles." Journal of Theoretical Biology, \textbf{82}, pp. 15-39 (1980).

\bibitem{bhamre2015} T. Bhamre, T. Zhang, and A. Singer. "Orthogonal Matrix Retrieval In Cryo-Electron Microscopy". IEEE International Symposium on Biomedical Imaging: from nano to macro (2015).



\bibitem{ppt} ``Signal Processing and Representation Theory," Class slides of the Johns Hopkins University.  


\bibitem{xian2018} Y. Xian, H. Gu, W. Wang, X. Huang, Y. Yao, Y. Wang, and J-Feng. ``Data-Driven Tight Frame for Cryo-EM Image Denoising and Conformational Classification." In 2018 IEEE Global Conference on Signal and Information Processing (GlobalSIP), pp. 544-548. IEEE, 2018.

\bibitem{xian2017} Y. Xian, Y. Pu, Z. Gan, L. Lu and A. Thompson. ``Adaptive DCTNet for audio signal classification". IEEE International Conference on Acoustics, Speech and Signal Processing (ICASSP), pp. 3999-4003, 2017.

\bibitem{xian2015} Y. Xian, A. Thompson, Q. Qiu, L. Nolte, D. Nowacek, J. Lu, and R. Calderbank, R. ``Classification of whale vocalizations using the weyl transform." IEEE International Conference on Acoustics, Speech and Signal Processing (ICASSP), pp. 773-777, 2015.

\end{thebibliography}
\end{document}